\documentclass[11pt]{article}
\usepackage{graphicx}
\usepackage{epsfig}

\newcommand{\BABARPubYear}    {06}
\newcommand{\BABARPubNumber}  {094}
\newcommand{\SLACPubNumber} {12148}

\input pubboard/babarsym

\newcommand{\runfive}{\ensuremath{316~\invfb}\xspace}
 
\long\def\inst#1{\par\nobreak\kern 4pt\nobreak
    {\it #1}\par\vskip 10pt plus 3pt minus 3pt}

\begin{document}
{\pagestyle{empty}

\begin{flushright}
SLAC-PUB-\SLACPubNumber \\
\babar-PROC-\BABARPubYear/\BABARPubNumber \\
\end{flushright}

\par\vskip 5cm

\begin{center}
\Large \bf Measurement of Direct CP Asymmetries \\
in Charmless Hadronic B Decays
\end{center}
\bigskip

\begin{center}
Emanuele Di Marco \\
Universit\`{a} di Roma ``La Sapienza''\\ 
Physics Department, P.le Aldo Moro 2, 00185 Roma, Italy\\
(representing the \babar\ Collaboration)
\mbox{ }\\
\today
\end{center}
\bigskip \bigskip

\begin{center}
\large \bf Abstract
\end{center}
We present recent results on time integrated and time dependent \CP violation 
for charmless hadronic $B$ decays using \babar{} detector at the \pep2{} B-factory.
\vfill
\begin{center}

Submitted to the the Proceedings of 33$^{\rm rd}$ International Conference on 
High-Energy Physics, ICHEP 06,\\
26 July---2 August 2006, Moscow, Russia.

\end{center}

\vspace{1.0cm}
\begin{center}
{\em Stanford Linear Accelerator Center, Stanford University, 
Stanford, CA 94309} \\ \vspace{0.1cm}\hrule\vspace{0.1cm}
Work supported in part by Department of Energy contract DE-AC02-76SF00515.
\end{center}

\newpage
} 

%
%


\section{Introduction}
\CP violation has been established in processes involving \Bz-\Bzb oscillations through 
measurements of the time dependence of neutral $B$-meson decays to final states which include
charmonium~\cite{Aubert:2006aq}.
Direct \CP violation occurs when the amplitude for a process $i \to f$ is different from
the \CP conjugate one: $\bar i \to \bar f$.
In the Standard Model (SM) this can occur as a result of interference among 
at least two contributing amplitudes which carry different weak and strong phases.
For a decay process $B \to f$ and its \CP conjugate $\bar B \to \bar f$ the direct \CP
asymmetry is defined as:
\begin{equation}
{\cal A}_{CP}=\frac{\Gamma(\bar B \to \bar f)-\Gamma(B \to f)}
{\Gamma(\bar B \to \bar f)+\Gamma(B \to f)}
\end{equation}
where $B$ can be either a charged or neutral $B$ meson.
If $A_1$ and $A_2$ are two of the contributing amplitudes to the decay, which have 
$\Delta\phi_w$ weak phase difference and $\Delta\phi_s$ strong phase difference, 
the direct \CP asymmetry can be expressed by:
\begin{equation} 
{\cal A}_{CP}=\frac {2\sin\Delta\phi_w \sin\Delta\phi_s}
{\vert A_1/A_2\vert + \vert A_2/A_1\vert + 2\cos \Delta\phi_w\Delta\phi_s}.
\end{equation}
This implies that large direct \CP violation is expected when the contributing amplitudes
have large $\Delta\phi_s$, $\Delta\phi_w$ and similar magnitudes. 
This happens in charmless $B$ decays where there are often contributions from Cabibbo 
suppressed tree amplitudes as well as from penguin diagrams which have different phases
but similar magnitudes. The best candidate in this sense is $\Bz\to\Kp\pim$ decay.

The simplest approach to the direct \CP violation measurement is to measure time 
integrated asymmetries, for example different rates for the process $B^+ \to f^+$ versus 
$B^- \to f^-$, or in self tagging neutral $B$ decays, for which the final state particles 
indicates the flavour of the $B$.

Direct \CP violation can also be observed as a time dependent asymmetry. In neutral $B$ 
decays to a \CP eigenstate the time dependent asymmetry can be expressed by:
\begin{equation}
{\cal A}_{CP}(\deltat)=S \sin(\deltamd\deltat) + C \cos(\deltamd\deltat)
\end{equation}
where \deltamd is the \Bz-\Bzb mixing frequency and $\deltat$ is the difference between the 
signal and tag $B$ decay times.
If the parameter $C$ is different from zero this is an indication of direct \CP violation.
In this case we define the direct \CP violation ${\cal A}_{CP}=-C$.

\CP violation in charmless $B$ decays can be also interpreted in terms of parameters of the 
Unitarity Triangle which describes the unitarity of the Cabibbo-Kobayashi-Maskawa 
quark mixing matrix~\cite{UTfitters}.

In this note we briefly review some recent measurements of direct \CP violation at the 
\babar{} experiment situated at the \pep2{} \epem storage ring at the Stanford Linear Accelerator
Center~\cite{Aubert:2001tu}. The analyses presented mostly uses about \runfive of data recorded at 
the \FourS resonance.

\section{Basic Measurement Strategy}
Events are fully reconstructed combining tracks and neutral clusters in the detector.
In order to select $B$ candidates we use a set of two kinematic variables:
the beam-energy-substituted mass 
$\mes=\sqrt{(s/2+{\bf p}_i \cdot {\bf p}_B)^2/E_i^2+p^2_B}$, 
and the energy difference $\DeltaE=E^*_B-\sqrt{s}/2$. 
Here, $(E_i,{\bf p}_i)$ is the
four-vector of the initial \epem{} system, $\sqrt{s} $
is the center-of-mass energy, ${\bf p}_B$ is the
reconstructed momentum of the \Bz{} candidate, and $E_B^*$ is its
energy calculated in the \epem{} rest frame. For signal decays,
the \mes{} distribution peaks near the \Bz{} mass with a
resolution of about $2.5$ \mevcc, and the \DeltaE{}
distribution peaks near zero with a resolution of  $10-50$ \mev,
depending on the final state.
\DeltaE depends on the mass hypothesis assigned to the tracks and
is so used in final states with kaons or pions. 
Particle identification (PID) also uses the the Cherenkov detector
(DIRC), which gives excellent $K$/$\pi$ separation ($>2.5 \sigma$ for
momenta $<4.0$ \gevc).

The main background comes from light quark continuum production,
which is different by signal in the more jet-like distribution of the 
decay products. Different variables, as the Legendre polynomials,
combined using algorithms which maximize the separation between signal
and background, 
are used to reject most of this background. Another source of 
background comes from decays of $B$ mesons which mimic the signal.
This background is typically more difficult to suppress.

The analysis strategy, common to the most of the analyses, is to perform
an unbinned maximum likelihood (ML) fit to several selection variables to 
extract \CP asymmetries simultaneously to the signal yields.
Large sidebands, {\it i.e.} regions where the signal contribution is negligible, 
are kept in the selection in order to characterize the background directly on data.
Background from other $B$ decays is usually estimated from Monte Carlo samples.

\section{Results on direct \CP Violation}
\boldmath
\subsection{$\Bz\to\Kp\pim$, $\Bz\to\pipi$}
\begin{figure}[htb]
  \begin{center}
    \epsfig{file=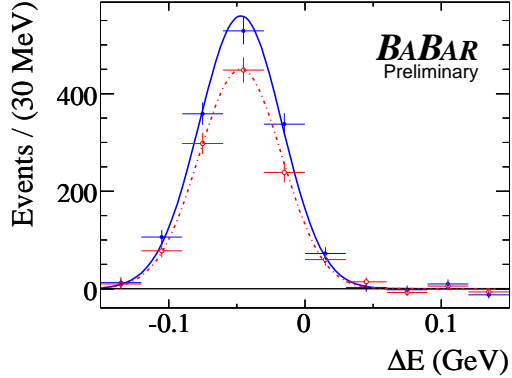,width=8.0cm}
  \end{center}
  \caption{The background-subtracted distribution of $\Delta E$ for signal $K^{\pm}\pi^{\mp}$
    events, comparing \Bz (solid) and \Bzb decays (dashed).
    \label{fig:Akpi}}
\end{figure}

\unboldmath
In $\Bz\to\Kp\pim$ decay the charge of the kaon can be used to infer the tag of the $B$ meson,
so a time integrated \CP violation measurement is feasible.
The analysis reconstructs simultaneously $\Bz\to h^+h^-$, with $h=K,\pi$, making use 
of \mes, \DeltaE, a Fisher discriminant $\cal F$ with event shape variables~\cite{fisher}
and the Cherenkov angle $\theta_c$.
\DeltaE and $\theta_c$ lead to the separation of the $h^+h^-$ modes.
The $\theta_c$ PDFs are determined separately for positive and negative tracks from
dedicated $D^{*+}\to \Dz\pip$, $\Dz\to\Km\pip$ data control samples.
The fit to 347 million $B \bar B$ pairs returns 2542 $\pm$ 67 $\Bz\to K\pi$ events,
and an asymmetry ${\cal A}_{K\pi}=-0.108 \pm 0.024 \pm 0.008$. As a cross-check, the measured
\CP asymmetry for the background is consistent with zero.
Fig.~\ref{fig:Akpi} shows the background-subtracted \DeltaE distribution for $\Kp\pim$
events and $\Km\pip$ events. The significance of the \CP violation, including systematics,
is 4.3$\sigma$~\cite{unknown:2006ap}.

The \CP asymmetry for $\Bz\to\pipi$ is measured with a time-dependent fit to the
$\Bz\to h^+h^-$ sample. The fit yields $675 \pm 42$ signal events, with a direct \CP
asymmetry parameter ${\cal A}_{CP}=-C=0.16 \pm 0.11 \pm 0.03$, consistent with zero.

\subsection{$\Bp\to h^+\piz$}
In the SM the charge asymmetry in $\Bp\to\Kp\piz$ is expected to be of the same order
of the $\Bz\to\Kp\pim$ one, while no asymmetry is expected in $\Bp\to\pi^+\piz$,
since the decay is mediated by only one amplitude. The main $B$ backgrounds for these 
modes are $B \to \rho \pi$, $B \to \rho K$ and $B \to K^* \pi$ decays.
The fit strategy is similar to the one of $\Bz\to\Kp\pim$.
The measured yields, branching fractions and \CP asymmetries are shown in 
Table~\ref{tab:hpiz}. The asymmetries are consistent with zero~\cite{unknown:2006ap}.
\begin{table}
  \begin{center}
    {\begin{tabular}{@{}lcc@{}}
        \hline\hline
        & $\Bp\to\pip\piz$                 &  $\Bp\to\Kp\piz$  \\
        \hline
       $N_{\mathrm{S}}$             & $572\pm 53$                      & $1239\pm 52$                 \\
       \BR($10^{-6}$)                        & $5.12\pm 0.47\pm 0.29$           & $13.3\pm 0.56\pm 0.64$       \\
       ${\cal A}_{CP}$                   & $-0.02 \pm 0.09 \pm 0.01$        & $0.02 \pm 0.04 \pm 0.01$  \\
       \hline\hline
     \end{tabular}}
    \caption{Signal yields ($N_{\mathrm{S}}$), branching fractions (\BR) 
      and \CP asymmetries for $\Bp\to h^+\piz$. \label{tab:hpiz}}
 \end{center}
\end{table}
\subsection{$\Bp\to\phi\Kp$, $B\to\phi\pi$}
The $\Bp\to\phi\Kp$ and $B^{\pm,0}\to \phi \pi^{\pm,0}$ in the SM proceed only through
$b\to s$ and $b\to d$ gluonic penguins, respectively, then the direct \CP asymmetry is 
expected to be zero. However, these decays are sensitive to new physics entering the loop
amplitude and providing additional phases, which could provide large \CP asymmetries and 
a $\BR(B\to\phi\pi)>10^{-8}$.
The analysis uses an extended ML fit to kinematic and event shape variables, the helicity of
the $\phi$ meson and $\theta_c$. The $\phi$ meson is defined as the $\Kp\Km$ pair 
with an invariant mass in $1.0045<m_{\Kp\Km}<1.0345$ \gevcc. 
The extracted yield for $\Bp\to \phi\Kp$ in 347 million $B\bar B$ pairs is 624 $\pm$ 30
events, with a direct \CP asymmetry ${\cal A}_{CP}=0.046 \pm 0.046 \pm 0.017$, which is 
consistent with zero~\cite{unknown:2006av}. 

A fit to 232 million $B \bar B$ pair returns no signal events for $B\to \phi \pi$ decays,
allowing to set upper limits on \BR{}'s: $\BR(B^+\to\phi\pi^+)<2.4\times 10^{-6}$ and
$\BR(B^0\to\phi\pi^0)<2.8\times 10^{-6}$~\cite{Aubert:2006nn}.

\subsection{$\Bp\to\Kz h^+$, $\Bz\to\bar\Kz\Kz$}
\begin{figure}[htb]
  \begin{center}
    \setlength{\tabcolsep}{0.1pc}
    \begin{tabular}{cc}
    \epsfig{file=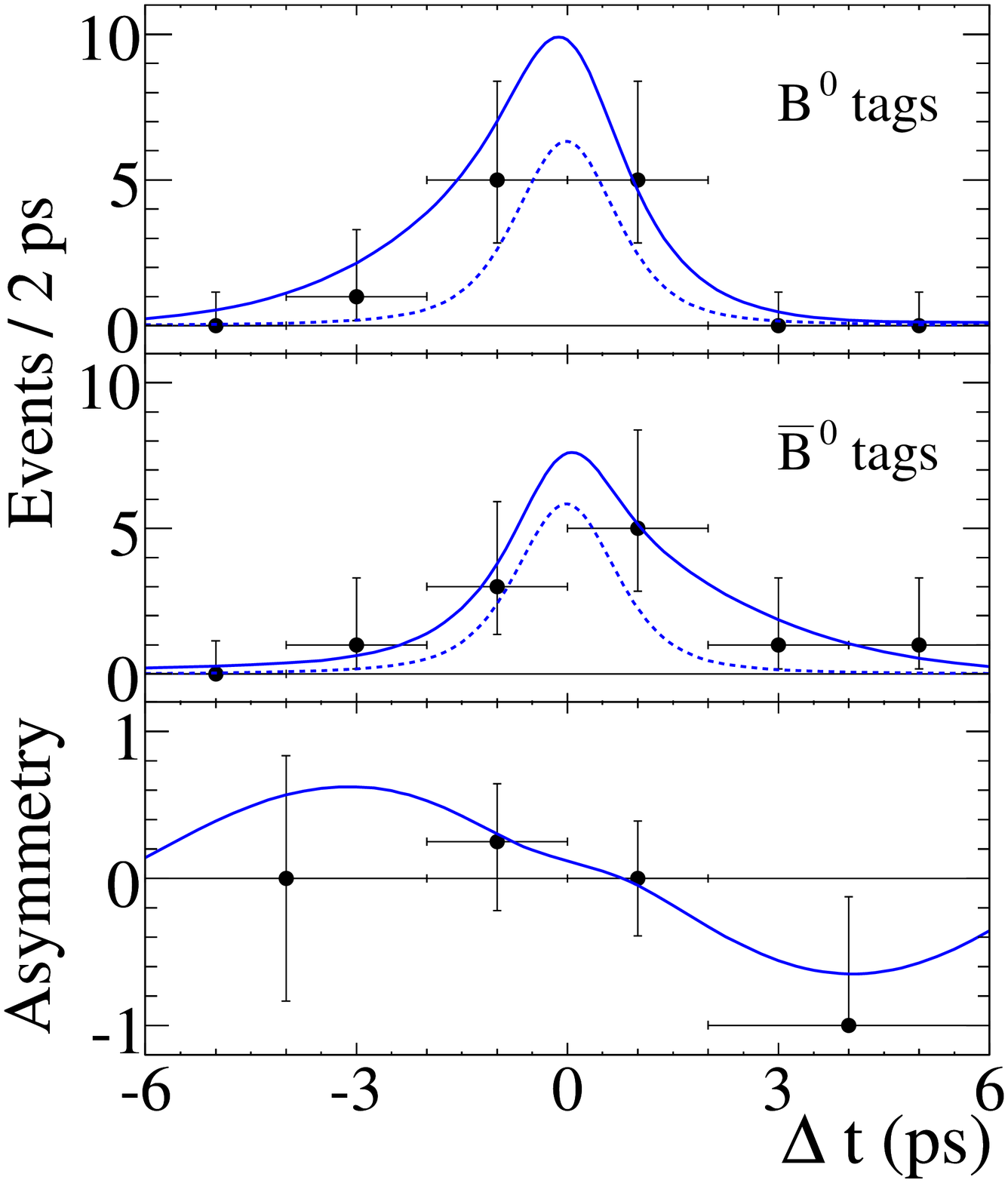,width=8.0cm,height=8.0cm} &
    \epsfig{file=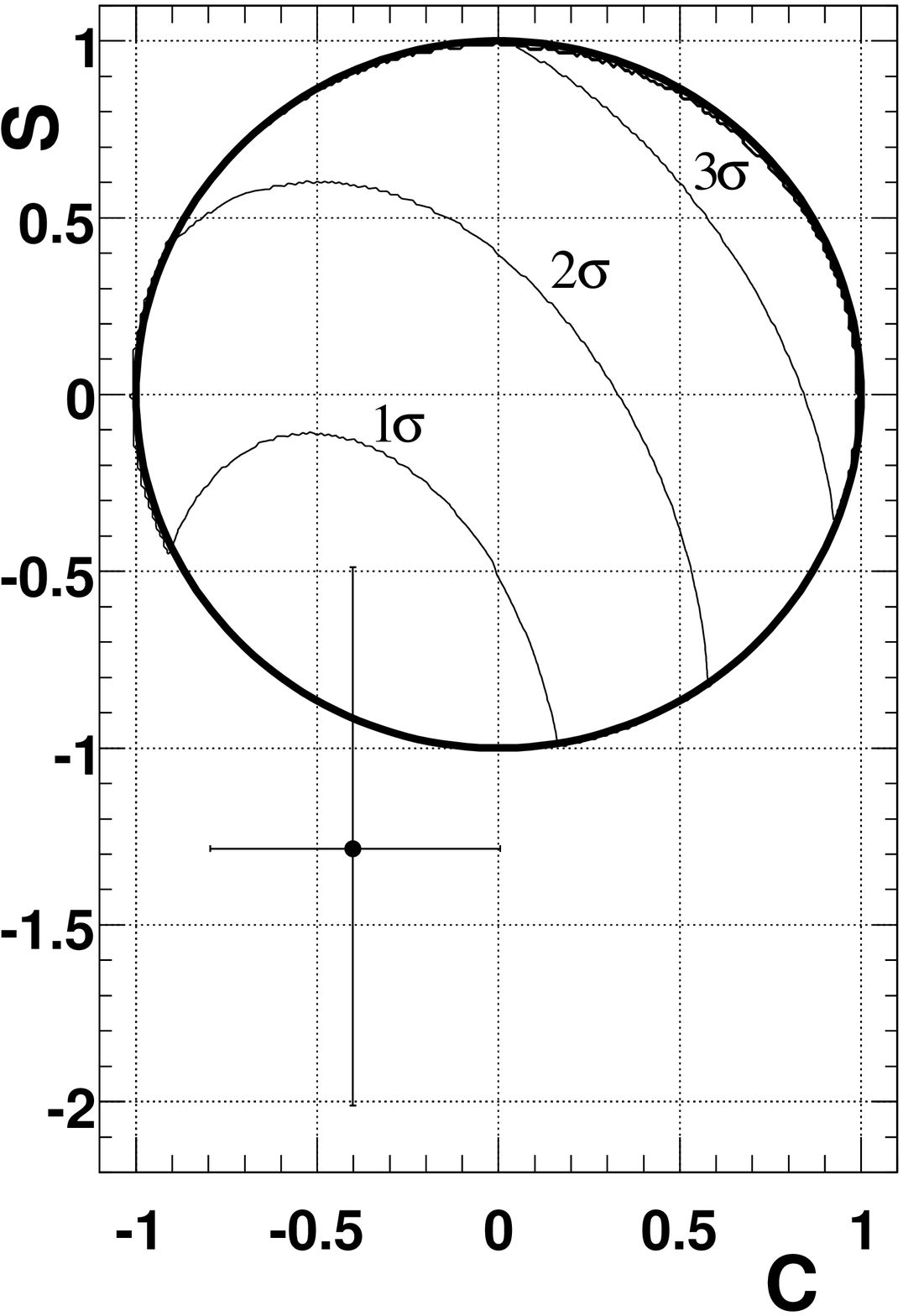,width=5.6cm,height=8.6cm} \\
    \end{tabular}
  \end{center}
  \caption{Left: \deltat distributions for $\Bz\to\KS\KS$ decays in data tagged as \Bz (top)
    or \Bzb (middle), and the asymmetry (bottom). 
    The solid (dotted) curve represents the total (background only) likelihood projection.
    Right: Confidence level contours in the $C$ vs. $S$ plane. The circle indicates the physically 
    allowed region, while the point with errors is the result on data.
    \label{fig:K0K0}}
\end{figure}

In SM, decays $\Bp\to\bar\Kz h^+$ and $\Bz\to\Kz\bar\Kz$ proceed only through penguin
transition $b\to d$, then the direct \CP asymmetry is expected to vanish.
In a way analogous to $\Bp\to\phi\Kp$, new physics effects could enhance it.
The direct \CP asymmetry in the neutral mode is measured with a time-dependent fit.
From an extended ML fit to \mes, \DeltaE and $\cal F$ (including \deltat for $\Bz\to\bar\Kz \Kz$), 
in a sample of 347 million $B\bar B$ pairs,
the signal yields and the \CP asymmetries are extracted. The results are reported in 
Table~\ref{tab:KK}. We observe for the first time the decays 
$\Bp\to\bar\Kz\Kp$ and $\Bz\to\bar\Kz\Kz$ with
significances of 5.3$\sigma$ and 7.3$\sigma$, respectively. The measured direct \CP asymmetries
are consistent with zero~\cite{:2006gm}, in agreement with theoretical expectations.
In Fig.~\ref{fig:K0K0} the asymmetry for $\KS\KS$ events and 
the confidence level contours in the $C$ vs. $S$ plane are shown.
\begin{table}
\begin{center}
    {\begin{tabular}{@{}lcc@{}}
      \hline\hline
      Mode & \BR ($10^{-6}$) & ${\cal A}_{CP}$  \\
      \hline
      $\Bp\to\Kz\pip$    & $23.9 \pm 1.1 \pm 1.0$   & $-0.03 \pm 0.04 \pm 0.01$ \\
      $\Bp\to\bar\Kz\Kp$ & $1.6 \pm 0.4 \pm 0.1$    & $0.10 \pm 0.26 \pm 0.03$ \\ 
      $\Bz\to\Kz\bar\Kz$ & $1.1 \pm 0.3 \pm 0.1$    & $0.40 \pm 0.41 \pm 0.06$ \\
      \hline\hline
    \end{tabular}}
  \caption{Signal yields ($N_{\mathrm{S}}$), branching fractions \BR ($\times 10^{-6}$) 
    and \CP asymmetries for $\Bp\to\Kz h^+$, $\Bp\to\Kz\bar\Kz$. \label{tab:KK}}
  \end{center}
\end{table}

\subsection{$\Bz\to(\rho\pi)^0$}
\begin{figure}[htb]
  \begin{center}
    \epsfig{file=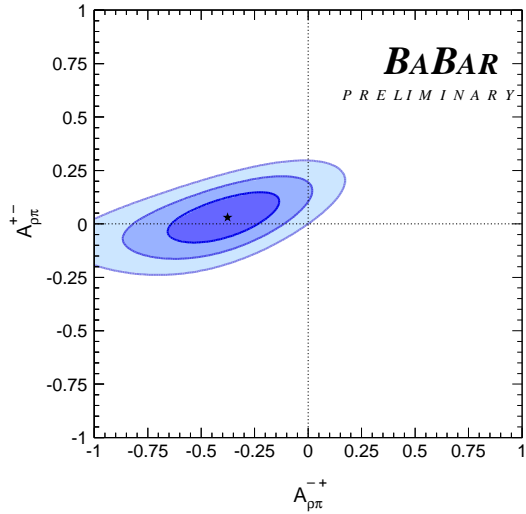,width=7cm}
  \end{center}
  \caption{Confidence level contours for the direct \CP asymmetries ${\cal A}_{\rho\pi}^{+-}$ and
    ${\cal A}_{\rho\pi}^{-+}$ . The shaded areas represent 1$\sigma$, 2$\sigma$ and 3$\sigma$ 
    contours, from darker to lighter.
    \label{fig:rhopi}}
\end{figure}

A time-dependent analysis of the Dalitz plot of the $\Bz\to\pip\pim\piz$ decays is performed,
in order to measure the \CP violation taking into account the interference
between the intermediate states: $\Bz\to\rho^{\pm}\pi^{\mp}$ and 
the color suppressed $\Bz\to\rho^0\piz$. 
QCD factorization predicts null direct \CP violation, due to the lack of penguin contributions,
but non-factorizable effects could be present, modifying this conclusion.
Through the interference between these amplitudes this measurement can lead to an 
unambiguous determination of the angle $\alpha$ of the Unitarity Triangle~\cite{Snyder:1993mx}.
An extended ML fit to \mes, \DeltaE, $\cal F$, Dalitz plot variables and $\deltat$
is performed to 347 million $B\bar B$ pairs.
Since the final states are not \CP eigenstates the \CP violation can be expressed 
in terms of the time-dependent $C$ parameter and time- and flavour-integrated \CP
asymmetry ${\cal A}_{\rho\pi}$. These can be rearranged in terms of ${\cal A}_{\rho\pi}^{+-}$ which 
describes the direct \CP violation of the amplitude where the $\rho$ is emitted from 
the $W$, and ${\cal A}_{\rho\pi}^{-+}$ which describes amplitudes where the $\pi$ is emitted from the $W$.
The fit returns a signal yield of $1847 \pm 69$ events,
dominated by $\Bz\to\rho^\pm\pi^\mp$ decays, and direct \CP asymmetries 
${\cal A}_{\rho\pi}^{+-}=0.03 \pm 0.07 \pm 0.03$,
${\cal A}_{\rho\pi}^{-+}=-0.38^{+0.15}_{-0.16} \pm 0.07$.
These results show an evidence of direct \CP violation, including systematics, of 3.0$\sigma$
~\cite{Aubert:2006fg}.
Fig.~\ref{fig:rhopi} shows confidence level contours for ${\cal A}_{\rho\pi}^{+-}$ and ${\cal A}_{\rho\pi}^{-+}$.

\section{Conclusions}
The direct \CP violation has been measured in $\Bz\to\Kp\pim$ decays to be of 0.108 $\pm$ 0.025,
resulting in a 4.3$\sigma$ effect.
We observed for the first time the decays $\Bp\to\bar\Kz\Kp$ and $\Bz\to\Kz\bar\Kz$, 
with  no \CP violation.
The direct \CP asymmetry is consistent with zero in other rare decay modes, 
consistently with the SM predictions for those decays which are 
mostly sensitive to new physics effects, as $\Bp\to\phi\Kp$.
In the time-dependent analysis of $\Bz\to(\rho\pi)^0$ we see an evidence of direct \CP violation
with 3.0$\sigma$ significance.
The increasing of the recorded luminosity in the next years will provide a powerful
tool to investigate the SM picture of the nature of \CP violation.

\section{Acknowledgements}
\label{sec:Acknowledgments}

I am grateful to my \babar\ collegues Gianluca Cavoto, Maurizio Pierini,
Denis Dujmic, and Mahalaxmi Krishnamurthy for their support in this work.
I would like to thank also Daniele del Re and prof. Aaron Roodman for their
support during the conference.  

\end{document}